\documentclass{cargese}\usepackage{graphicx}
\let\footnote\savefootnote
\let\footnotetext\savefootnotetext 
\let\footnoterule\savefootnoterule 
\normallatexbib
\font\mybb=msbm10 at 12pt
\font\mybbsub=msbm10 at 10pt
\font\myeu=eufm10 at 12pt
\def\bb#1{\hbox{\mybb#1}}
\def\bbsub#1{\hbox{\mybbsub#1}}
\def\frak#1{\hbox{\myeu#1}}
\def\ZZ {\bb{Z}}

\def\ZZsub {\bbsub{Z}}

\def\beq{\begin{equation}}
\def\eeq{\end{equation}}
\def\bdm{\begin{displaymath}}
\def\edm{\end{displaymath}}
\def\beqa{\begin{eqnarray}}
\def\eeqa{\end{eqnarray}}

\def\noi{\noindent}

\begin{document}


\articletitle[Discrete U-duality]{Discrete U-duality groups}


\author{Germar Schr\"oder}


\affil{Max-Planck-Institut f\"ur Gravitationsphysik,
Albert-Einstein-Institut,\\
Am M\"uhlenberg 1, D-14476 Golm, Germany}         
%
\email{germar@aei-potsdam.mpg.de}
%

\begin{abstract}
Generators for the discrete U-duality groups of toroidally 
compactified M-theory   
in $d\geq 4$ are presented and used to determine the $d=3$ U-duality group.
This contribution summarizes the results of
\cite{Miz99}.
\end{abstract}


\section{Introduction and Summary}

The study of nonperturbative duality symmetries in the last years has 
dramatically changed our understanding of string theory.
U-duality\index{U-duality},  
introduced in \cite{Hul95}, is one of these symmetries. 
It was conjectured to be generated by the
target-space duality of $\mbox{T}^d$ 
(see \cite{Giv94} for a review)
and the modular group
of the torus including the eleventh M-theory direction 
\cite{Sch95}.
This definition was 
adopted in the algebraic approach to U-duality reviewed in \cite{Obe98}.

Using instead the original conjecture of
\cite{Hul95} that takes the hidden symmetries of low energy 
supergravity as a starting point,
generators of discrete U-duality in four dimensions
may be determined directly. Groups in higher 
dimensions can be found by embeddings, 
and a method to determine the $d=3$ U-duality is presented.
Applied to a toy model corresponding to a truncation of M-theory, 
the method is seen to give a significantly different result than for 
the full theory.

\section{U-duality in Four Dimensions}

Compactifying eleven dimensional supergravity, 
the low energy limit of M-theory, on
$\mbox{T}^7$ yields 70 scalars arising from the 
$d=11$ 3-form potential and 
the moduli of the torus. They may be joint in a field 
${\cal V}^{(4)}
\in E_{7(+7)}/SU(8)$ \cite{Cre79}, a representation matrix 
in the fundamental {\bf 56} representation.
Furthermore, 28 $U(1)$ gauge fields arise from the 
$d=11$ potential and metric. These fields and their magnetic duals 
can be arranged to 28 dimensional vectors 
${\cal G}_{\bar{\mu}\bar{\nu}}$ and
${\cal H}_{\bar{\mu}\bar{\nu}}$, $\bar{\mu},\bar{\nu}=0\dots 3$.

The equations of motion of the theory are invariant under classical
$E_{7(+7)}$ duality and local $SU(8)$ transformations, acting as
\bdm
{\cal F}_{\bar{\mu}\bar{\nu}}\to\Lambda^{-1}
{\cal F}_{\bar{\mu}\bar{\nu}},\ \
{\cal V}^{(4)}\rightarrow h\; {\cal V}^{(4)} \Lambda,
\ \ \Lambda \in E_{7(+7)},\ \ h \in SU(8)
\edm
where ${\cal F}^t = ( {\cal G}_{\bar{\mu}\bar{\nu}}, 
{\cal H}_{\bar{\mu}\bar{\nu}})^t$.
Defining charges ${\cal Z}= \oint_\Sigma {\cal F}$, 
the DSZ quantization condition breaks $E_{7(+7)}$ to a 
discrete subgroup inducing integer 
shifts on the charge lattice. This group is the U-duality group 
and was
proposed to extend to a nonperturbative quantum 
symmetry of M-theory.

To make U-duality transformations ``manageable'' in $d=4$, 
the {\bf 56} representation needs to be addressed. This can be done 
by an embedding into $\frak{e}_{8(+8)}$ using Freudenthal's
realization of exceptional Lie algebras \cite{Freudenthal}.  
\index{exceptional Lie algebras, Freudenthal's realization}
The $\frak{e}_{8(+8)}$ generators are given by
$E^i_{\ j},\ \ i,j = 1\dots 9$, corresponding to $\frak{sl}_9$, and
$E^{ijk}$, $E^*_{ijk}$. 
Their commutators are given in \cite{Miz99}.
Defining the basis
\beqa
{\cal S}^t =\Big(-E^*_{\bar{i}\bar{j}9},
+E^1_{~\bar{i}}\; | \;
-E^{1\bar{i}\bar{j}}, - E^{\bar{i}}_{~9}\Big),\ \ 
{\cal X}^t =\Big( x^{\bar{i}\bar{j}}, x^{\bar{i}\bar{9}} 
\; | \;
 y_{\bar{i}\bar{j}}, y_{\bar{i}\bar{9}} \Big),
\label{rep2}
\eeqa
$\bar{i},\bar{j}=2 \dots 8$,
the adjoint action of the $\frak{e}_{7(+7)}$ 
subalgebra on ${\cal X}{\cal S}$ exactly spans the
{\bf 56} representation as defined in \cite{Cre79}. 
The basis (\ref{rep2}) can therefore
be used to study U-duality transformations on ${\cal F}$ and 
${\cal V}^{(4)}$, as well as its subgroups
T-duality and S-duality.

What are the generators of $E_{7(+7)}(\ZZ)$? Using the fact that
the ${\bf 56}$ representation of $E_{7(+7)}$
is the unique minimal representation of $E_{7(+7)}$, 
it may be proven using the Birkhoff 
decomposition of Lie groups 
that the subgroup of $E_{7(+7)}$ inducing integer shifts
on the lattice defined by ${\cal S}$
is generated by ``fundamental unipotents''\footnote{
See \cite{Matsumato}, where this group is defined in a more general context
as homeomorphism of the group ring over $\ZZsub$ to $\ZZsub$.}, that is, the 
action of the discrete subgroup is spanned by exponentiating the Chevalley 
generators for all positive and negative roots. 
\index{discrete exceptional U-duality groups} From these, 
$T$ and $S$ generators may be built parallel to $SL(2,\ZZ)$, 
the latter carrying a representation of the
Weyl group modulo $\ZZ_2$. This together with the basis ${\cal S}$ 
yields contact to the algebraic approach to M-theory, and it may be 
shown that the two approaches are equivalent. 

Since the notion of the above generators is representation 
independent, the discrete U-duality groups of higher dimensional 
theories follow directly from truncating the Dynkin diagram.
Their representations are minimal and can be read off from ${\cal S}$.

\subsection{U-duality in Three Dimensions}

The $d=3$ theory is known to have classical $E_{8(+8)}$ symmetry.
As only scalars remain in the
theory, the notion of electric charge seems ill defined and the meaning of a 
duality symmetry seems unclear. \index{U-duality in d=3}
Therefore, in order to define U-duality in $d=3$, 
a method proposed in \cite{Hul95} parallel to \cite{Sen95} 
may be extended  to M-theory.
By compactifying
M-theory on the torus, we can choose eight 
different ways how to
compactify first to four dimensions. This
results in eight $E_{7(+7)}(\ZZ)$'s acting {\it differently} 
on M-theory
fields. By reducing the theory further to three dimensions, 
these
groups are merged together to form the three dimensional 
duality group.

The reduction to $d=3$ yields a scalar coset matrix of the form
\beq
{\cal V}^{(3)} = 
{\cal V}'^{(4)}
\exp\Big(\frac12\phi\ \sum_{i=1}^8 h_i \Big)
\exp\Big({{\cal Y}\cdot {\cal S}}\Big)
\exp\Big({f\ E^1_{\ 9}}\Big).
\label{3de7}
\eeq
${\cal V}'^{(4)}$ is identical to ${\cal V}^{(4)}$,  but now in 
the {\bf 248} adjoint representation of $E_{8(+8)}$. ${\cal Y}$ obeys
$\partial_{\mu} \eta = {\cal G}_{\mu z},\ \ 
\partial_{\mu} \bar{\eta} = {\cal H}_{\mu z}, \ \ 
{\cal Y}= ( \eta, \bar{\eta})^t,\ \ \mu,\nu=0\dots 2$ 
and therefore carries the 
$d=4$ charges, where $z$ is the compact fourth direction. 
$\varphi$ and $f$ are the $d=3$ dilaton and dualized KK field strength
respectively, and $h_i$ represents the $\frak{sl}_9$ Cartan subalgebra. 

A $\Lambda\in E_{7(+7)}\subset E_{8(+8)}$ transformation
acts on ${\cal V}'^{(4)}$ and ${\cal Y}$ exactly as discussed in the last 
section. It therefore represents the $d=4$ U-duality in $d=3$. 
Completing a circle around vortex solutions  
in $d=3$ may be seen to correspond 
to an $E_{7(+7)}(\ZZ)$ action on the fields,
parallel to \cite{Sen95}.\index{vortex solutions}
  
The different compactifications 
yield 8 different coset matrices in $d=3$. Using  
explicitly the connection to M-theory fields, it can be seen that 
they are related by
\bdm
{\cal V}^{(3)}_{\# n} = 
(P_n S^1_{\ n})^{-1} \; h_n\;\ 
{\cal V}^{(3)}_{\# 1}\;\ P_n S^1_{\ n} 
\edm
where $h_n$ is a local transformation restoring upper triangular 
paramete\-ri\-zation, $S^1_{\ n}$ corresponds to a Weyl reflection 
and $P_n$ corresponds to a charge conjugation in
$d=4$. 
Denoting the $d=4$ 
U-duality generators of the $n$th compactification by $\Lambda$,
the total $d=3$ U-duality is given by joining all generators

\bdm
P_n S^1_{\ n} \ \Lambda\  (P_n S^1_{\ n})^{-1}
\edm

\noi  
for all $n$. This can be seen to yield the whole 
$E_{8(+8)}(\ZZ)$ defined by exponentiating all Chevalley generators.
The intersection of two different $d=4$  U-dualities is seen to be 
$E_{6(+6)}(\ZZ)$ as expected.
This determines U-duality in $d=3$.

\subsection{$G_{2(+2)}$ as Toy Model}

The described method to generate 
$d=3$ U-duality can also be applied to 
five dimensional simple supergravity
as toy model, 
which upon
reduction to three dimensions exhibits a $G_{2(+2)}$ 
global symmetry
\cite{Miz98}. 
\index{$G_{2(+2)}$ supergravity in d=3}
It is known that this theory closely resembles $d=11$ 
supergravity in many respects and corresponds to a direct truncation.
The analogue of U-duality in $d=4$ is $SL(2,\ZZ)$, acting in the 
spin 3/2 representation.

Here, the joint U-duality $U(\ZZ)$ in $d=3$ is strictly smaller 
than $G_{2(+2)}(\ZZ)$
defined by exponentiating the Chevalley generators for all roots. All 
generators of $U(\ZZ)$ correspond to short roots of $G_{2(+2)}(\ZZ)$. 
That the groups do not agree is therefore connected to the fact that 
$G_{2(+2)}$ is not simply laced. Since no string 
compactification described by this no-moduli supergravity at low 
energies is known, one cannot determine which group is the ``correct'' 
U-duality group until such a description has been found. 


\begin{acknowledgments}
I wish to thank Shun'ya Mizoguchi for a stimulating and fruitful 
collaboration.
\end{acknowledgments}



%

\end{document}